\documentclass[a4paper,11pt]{article}
\pdfoutput=1 

\usepackage{jheppub} 

\usepackage[T1]{fontenc} 

\title{\boldmath The Schwarzschild-Black String AdS Soliton: Instability and Holographic Heat Transport}

 \author{Felix M.\ Haehl}
 \affiliation{Institute for Theoretical Physics\\
     ETH Zurich\\ CH-8093 Zurich\\ Switzerland}

\emailAdd{haehlf@student.ethz.ch}

\def\bp{\begin{pmatrix}}
\def\ep{\end{pmatrix}}
\def\ba{\begin{align}}
\def\ea{\end{align}}

\def\l({\left(}
\def\r){\right)}

\def\be{\begin{equation}}
\def\ee{\end{equation}}

\abstract{We present a calculation of two-point correlation functions of the
stress-energy tensor in the strongly-coupled, confining gauge theory which is holographically dual to the
AdS soliton geometry. 
The fact that the AdS soliton smoothly caps off at a certain point along
the holographic direction, ensures that these correlators are dominated by quasinormal mode contributions and 
thus show an exponential decay in position space. In order to study such a field theory on 
a curved spacetime, we foliate the six-dimensional 
AdS soliton with a Schwarzschild black hole. 
Via gauge/gravity duality, this new geometry describes a confining field theory with
supersymmetry breaking boundary conditions on a non-dynamical Schwarzschild black hole background.
We also calculate stress-energy correlators for this setting, thus demonstrating exponentially damped heat transport.
This analysis is valid in the confined phase. We model a deconfinement transition by explicitly demonstrating
a classical instability of Gregory-Laflamme-type of this bulk spacetime. 
}

\begin{document} 
\maketitle
\flushbottom

\section{Introduction}
The AdS/CFT conjecture \cite{AdSCFTmaldacena,AdSCFTwitten,AdSCFTGubser} has become a main tool to study
strongly coupled gauge theories from the point of view of a dual description in terms
of a suitable low energy limit of string theory. 
In particular, the dictionary of gauge/gravity duality provides a way to calculate
correlation functions of field theory operators from the gravitational dynamics of the bulk. 
The stress-energy tensor of the 
field theory is induced by the asymptotic behavior of the bulk metric itself: The propagation of small 
perturbations of the quantum stress tensor
can be studied by solving the problem of graviton propagation in the bulk spacetime. 
See \cite{PolicastroSonStarinets} for the calculation of stress-energy tensor correlation functions in
strongly coupled $\mathcal{N}=4$ supersymmetric Yang-Mills (SYM) theory, and \cite{MinkPrescr, QNMandHolography}
for the generalization to thermal $\mathcal{N}=4$ SYM theory. The poles of these retarded Green's functions 
are usually easier to calculate because they are just the quasinormal modes in the language of bulk gravity
\cite{QNMandHolography,BirminghamEtAl}.


Besides understanding supersymmetric and conformal field theories (in particular $\mathcal{N}=4$ SYM), 
one would like to describe strongly coupled gauge theories that share more properties with (large $N$) QCD. 
We will use an approach where we address the issue of finding bulk geometries which are appropriate 
to break supersymmetry and conformal invariance. 
Supersymmetry may be broken by considering a bulk which contains a Scherk-Schwarz compactified dimension 
and imposing antiperiodic boundary conditions on the fermions \cite{WittenProp}. 
It has been proposed to consider the AdS soliton as a bulk which
is only \textit{locally} asymptotically AdS and has a compact dimension (an $S^1$ circle)
\cite{WKBsoliton,AdSSoliton}. The introduction of 
a scale (the size of the compact dimension) also allows to break conformal invariance. 
The AdS soliton is constructed by a double analytic continuation of the 
planar AdS black hole such that a compactification takes place as one makes the original time coordinate 
Euclidean. This geometry is horizon-free and it may be visualized as the surface of one half of a cigar which caps off smoothly at its tip (\emph{infrared floor}). 
The entropy density therefore vanishes in the classical limit and the dual field
theory is in a confined phase. See \cite{AdSCFT_Mateos} for a discussion along
these lines, and \cite{AdSCFTPetersen} for an approach towards confinement. 
The AdS soliton is perturbatively stable and has the lowest 
energy in its class of spacetimes with the particular locally asymptotically AdS boundary conditions.

One expects
that the finite size of the geometry along the holographic direction induces a quantization of 
the graviton modes. These quasinormal frequencies appear as the poles of the retarded Green's function
in the dual quantum theory, as one would expect on general grounds \cite{QNMandHolography}; see also
\cite{HorowitzHubeny,MatchingSol}. We will explicitly calculate field theory correlators which 
describe momentum diffusion and we will see how the quantization of graviton modes leads to 
an exponential damping of such transport phenomena. 

After understanding some of these properties of the field theory dual to the AdS soliton, one can ask
what happens if this theory lives on a non-trivial background spacetime. An interesting feature of 
the AdS soliton is that it can be foliated with arbitrary Ricci flat slices which take the role 
of the boundary geometry in holography. 
As one foliates the AdS soliton with a Schwarzschild black hole, one obtains a six-dimensional
spacetime with two independent physical scales: 
the Schwarzschild radius $r_s$ and the size of the compact dimension $L_\tau$.
This \emph{Schwarzschild-black string AdS soliton} (\emph{Schwarzschild soliton string} for short)
serves as a simple model of a bulk geometry that describes
a non-supersymmetric, strongly coupled plasma around a (non-dynamical) 
black hole background. The geometric parameter $r_s$ corresponds 
to the plasma temperature $T\propto r_s$. We will address the question of how this field theory 
propagates thermal excitations near the black hole horizon towards infinity. Analogous considerations
as in the case of a flat background lead us to expect an exponential damping. 

Besides our interest in stress tensor correlators on a black hole background, 
another closely related aspect of these field theories is 
their deconfinement transition as one lowers $T$ \cite{MarolfAdSString}. 
Holographically, this phase transition can be understood
in terms of a
Hawking-Page transition between the Schwarzschild soliton string 
and the planar AdS black hole as one varies $r_s$ at fixed $L_\tau$. In order to determine the 
transition temperature, we will calculate at which point the Schwarzschild soliton
string becomes unstable against small perturbations. 
This computation is closely related to the calculation of thermal Green's functions because
it also involves solving the bulk graviton equations of motion.
Qualitatively, one expects to find a Gregory-Laflamme-type
instability \cite{GL1}
when the black hole horizon is small compared to the compact dimension.

This paper is organized as follows. 
In section \ref{sec:AdS_Soliton}, we use the gauge/gravity duality to calculate a stress-energy tensor
two-point function in the strongly coupled, confining gauge theory which is dual to the AdS soliton. 
In section \ref{sec:Instability}, we introduce the Schwarzschild soliton string and confirm quantitatively 
that it is indeed unstable against
small perturbations with tensor modes of a decomposition of the spacetime 
with respect to the base manifold $\mathfrak{B}=\text{Schw}_4$ (i.e.\ the Schwarzschild black hole). 
The instability problem reduces to a combination of the well-known Gregory-Laflamme instability
and the propagation of a scalar in the AdS soliton.
Once we have identified the stable phase of the Schwarzschild soliton string, section \ref{sec:StringCorrelators}
will be concerned with
the study of stress-energy correlation functions 
in the strongly coupled field theory on a Schwarzschild black hole background which is
dual to the Schwarzschild soliton string in gravity. An important ingredient for this calculation 
will be the results from section \ref{sec:AdS_Soliton}. Due to the fact that the boundary metric is no
longer translationally invariant, we will need to generalize the Fourier decomposition that could
otherwise be used.
We conclude with some remarks in section \ref{sec:conclusion}. 
In appendix \ref{appendix:AdSSoliton}, details of the calculation of the AdS soliton quasinormal modes 
are outlined.
We show in appendix \ref{appendix:vectorscalar} that vector and scalar perturbations do not destabilize the Schwarzschild soliton string. 
\vfill

\section{Shear Diffusion in the AdS Soliton Dual}
\label{sec:AdS_Soliton}
In this section we review some properties of the AdS soliton geometry and its dual confining field theory.
We calculate the quasinormal modes (QNM) for a linearized perturbation of this geometry which propagates like a 
scalar field. The QNM contribution with longest wavelength dominates stress-energy correlators in the dual field theory.
Besides being interesting for their own sake, results of this analysis will be needed for
calculations in section \ref{sec:StringCorrelators}.
By viewing our quantum theory as a toy model for QCD, the poles of the Green's functions may
be interpreted as glueball masses. Such holographic computations of glueball spectra in QCD${}_3$ and
QCD${}_4$ have been carried out, see e.g.\ \cite{glueballs}. The authors of \cite{WKBsoliton} even made use
of the same AdS soliton geometry and calculated glueball masses with a WKB approach.
We will make a comparison with their results. 

There have also been proposals for models which are more driven by phenomenology.
In \cite{AdSQCD} the bulk is constructed as a stack of flat slices which are conformally 
rescaled in such a way as to reproduce known
QCD phenomenology. This model relies on similar computations involving scalar 
field propagation and QNMs as the model that we will be studying. In this sense
our calculation might be easily adopted to such more phenomenologically relevant scenarios.
Models based on D-branes are also able to reproduce within certain bounds the phenomenology of QCD.
In particular, the Sakai-Sugimoto ansatz models QCD holographically by investigating stacks of 
D-branes, see \cite{SakaiSugimoto} and references therein.

\subsection{The AdS Soliton}
The AdS soliton has been described by Horowitz and Myers in \cite{AdSSoliton}. 
To construct the AdS soliton, we start with the following AdS black hole 
solution to $d$-dimensional Einstein gravity with
cosmological constant $\Lambda<0$:
\begin{equation} 
ds^2 = \frac{r^2}{\ell^2} \left[ - \left( 1- \frac{r_0^{d-1}}{r^{d-1}} \right)dt^2 + (dx^i)^2 \right] 
   + \left( 1-  \frac{r_0^{d-1}}{r^{d-1}} \right)^{-1} \frac{\ell^2}{r^2} dr^2 , \label{SolitonStart}
\end{equation}
where $\ell=-(d-1)(d-2)/2\Lambda$ is the AdS${}_d$ radius, and $i=1,\ldots ,d-2$. 
If we now perform a double analytic continuation of this metric, i.e.\ $t\rightarrow i\tau$ and $x^{d-2} \rightarrow it$,
we obtain 
\be
ds^2 = \frac{r^2}{\ell^2} \left[ \eta_{\mu\nu}dx^\mu dx^\nu + \left( 1- \frac{r_0^{d-1}}{r^{d-1}} \right) d\tau^2 \right] 
   + \left( 1- \frac{r_0^{d-1}}{r^{d-1}} \right)^{-1} \frac{\ell^2}{r^2} dr^2, \label{AdS_Soliton}
\ee 
where $\eta_{\mu\nu}$ is the $(d-2)$-dimensional Minkowski metric, and $(x^\mu)=(t,x^1,\ldots,x^{d-3} )$. 
Note that the coordinate $\tau$ has to be periodically identified in a Kaluza-Klein spirit in order to avoid a
conical singularity at $r=r_0$, i.e.\ $\tau \sim \tau + 4\pi\ell^2/(d-1)r_0$. 
We anticipate already at this point the following nice feature of the geometry (\ref{AdS_Soliton}): 
The flat space metric $\eta_{\mu\nu}$ can be replaced by any Ricci flat manifold and (\ref{AdS_Soliton})
will still be a solution of Einstein gravity. Since this part of the metric corresponds to the
non-compact boundary dimensions, we will be able to make a transition to boundary theories on a curved background
(see section \ref{sec:Instability}).

Since the $\tau$-circle closes
smoothly at $r=r_0$, the geometry just ends there and the entire spacetime (\ref{AdS_Soliton}) is horizon-free
and everywhere smooth. At $r=r_0$ there is no singularity but an \textit{infrared floor} where the spacetime ends in a cigar 
shaped geometry. The AdS soliton has a translational symmetry along the $(d-2)$ Minkowski coordinates, and a
$U(1)$ symmetry along the compact dimension.
Due to the periodicity in $\tau$ the AdS soliton is only locally asymptotically AdS.

What does the dual field theory look like? First of all, the compact dimension allows for supersymmetry breaking
by means of imposing antiperiodic boundary conditions for the fermions along the $\tau$-circle.
This introduces a mass gap in the fermionic spectrum such that the massive excitations decouple in the low energy effective theory.
The compact dimension also breaks conformal invariance which can be qualitatively understood by the fact that 
the decoupling of massive fermion modes changes the $\beta$-function (see also \cite{AdSCFTPetersen}). 
Since the AdS soliton geometry is horizon-free, 
the entropy of the AdS soliton vanishes to first order in $N^2$ (i.e.\ in the classical limit), as one would
expect for a field theory in a confined phase. A first order confinement-deconfinement phase transition at
a certain temperature $T_\text{dec.}>0$ is expected to happen in the field theory \cite{AdSCFT_Mateos}. 
On the gravity side, this corresponds to a 
Hawking-Page transition between the AdS soliton and a Schwarzschild-AdS black hole \cite{HawkingPage}.

\subsection{Energy-Momentum Correlators: Analytic Approach} \label{sec:ads_soliton_calc}
We now want to use gauge/gravity duality to calculate energy-momentum 
two-point functions in the boundary field theory of the AdS soliton. 
This analysis can be done in some analogy to \cite{Viscosity, AdSCFT_to_hydro} since the boundary metric is flat and 
we can therefore use the same simplifications that are used to calculate correlators in thermal $\mathcal{N}=4$ SYM.

In case of the AdS soliton we can consider fluctuations of $\phi\equiv h^2_1$ 
that propagate in the $x^3$-direction. This will eventually allow us to 
calculate the field theory correlator $\langle T_2^1 T_2^1 \rangle$ holographically.
By the same reasoning as in the case of thermal $\mathcal{N}=4$ SYM \cite{Viscosity}, 
the remaining $O(2)$ 
symmetry of the background metric ensures that $\phi$ decouples and
satisfies a massless scalar wave equation in the bulk metric.
 Rescaling the coordinates 
as $z=r_0/r$ and $y=r_0\tau/\ell^2$, the AdS soliton metric (\ref{AdS_Soliton}) becomes
\be
ds^2 = \frac{\ell^2}{z^2} \left[ \alpha^2 ds_\text{Mink.}^2 + (1-z^{d-1}) dy^2 + (1-z^{d-1})^{-1} dz^2 \right],
  \label{solitonMetric}
\ee
where $\alpha \equiv r_0/\ell^2$ and $ds_\text{Mink.}^2$ is the line element of $(d-2)$-dimensional Minkowski space.
From the massless scalar wave equation in this metric, $ \partial_a ( \sqrt{-g} g^{ab} \partial_b \phi)= 0$, we find
\begin{align}
 \frac{1}{\alpha^2} \hat \Box \phi - \left[ (d-1) z^{d-2} + \frac{4}{z}(1-z^{d-1}) \right] \partial_z \phi
   + (1-z^{d-1}) \partial_z^2 \phi + \frac{1}{(1-z^{d-1})} \partial_y^2 \phi = 0, \label{AdS_Soliton_allg}
\end{align}
where $\hat \Box$ is the wave operator on $(d-2)$-dimensional Minkowski space. 

From now on, we concentrate on 
$d=6$, although generalizations are straightforward. The case $d=6$ describes the dual of a gauge theory in $3+1$ Minkowski 
spacetime (times $S^1$) and is therefore particularly interesting. We make the additional assumption that the solution is independent 
of $y$, i.e.\ homogeneous along the compact circle. This assumption is well justified for the long wavelength limit that we are mainly
interested in. We now make the Fourier ansatz 
\begin{align}
  \phi(t,x^3,z)= \int \frac{d\omega\; dq}{(2\pi)^2} \; \phi_{(0)}(q) \phi_q(z) e^{-i\omega t + iqx^3} 
 \qquad \text{with } \phi_q(0) = 1 
\end{align}
such that $\phi_{(0)}(q)$ is the Fourier transform of the boundary value $\phi(t,x^3, 0)$, i.e.\ we demand the 
normalization $\phi_q(z=0) = 1$.
Eq.\ (\ref{AdS_Soliton_allg}) thus yields the mode equation
\be
 \phi_q'' - \left[\frac{5z^4}{(1-z^5)} + \frac{4}{z}\right] \phi_q' - \frac{q^2}{\alpha^2} 
  \frac{1}{(1-z^5)} \phi_q = 0\, ,  \label{AdS_Soliton_Eq}
\ee
where $(q^\mu)=(\omega,0,0,q)$ in the zero frequency limit (i.e.\ $\omega=0$) 
such that $m^2=-q^2$ is the boundary mass.\footnote{Note that we will often call the eigenvalues
of Eq.\ (\ref{AdS_Soliton_Eq}) quasinormal frequencies despite the fact that they are actually 
momentum eigenvalues at zero frequency.} This equation has
the following fundamental power series solutions near $z=0$:
\begin{align}
 \phi_{q,1}^{(0)} &= 1 - \frac{q^2}{6\alpha^2} z^2 + \frac{q^4}{24 \alpha^4} z^4 + \ldots \equiv \sum_{n=0}^\infty a_n z^n \, ,
   \label{SolitonSol1} \\
 \phi_{q,2}^{(0)} &= z^5 \left( 1 + \frac{q^2}{14 \alpha^2} z^2 + \ldots \right) \equiv \sum_{n=5}^\infty b_n z^n \, ,
   \label{SolitonSol2}
\end{align}
where the recursion relations for the coefficients $a_n$ and $b_n$ can be found in appendix \ref{appendix:AdSSoliton}. 

It has been argued that in real time thermal AdS/CFT the incoming wave boundary condition at the horizon
should be used to single out a unique solution \cite{Viscosity}. 
However, in the case of the AdS soliton, the solution that we find near $z=1$ is 
not of the form of an incoming or outgoing wave. 
This is related to the fact that the AdS soliton does not have a horizon and we have to 
impose another boundary condition. As pointed out by Witten \cite{WittenProp}, an important condition that should be
imposed for any acceptable solution is the Neumann condition $d\phi_q/d\rho = 0$ at the IR floor $z=1$. Here, $\rho$ is the 
natural coordinate in which the ``tip'' of the metric at $z=1$ looks like the origin of polar 
coordinates\footnote{The
coordinates $(\rho,\varphi)$ in which the IR-floor $z=1$ of the $d$-dimensional AdS soliton (\ref{solitonMetric})
looks like the origin in polar coordinates, 
are given by $\rho^2 = a (1-z^{-1})$, $\varphi = b y$ with $a=4\ell^2/(d-1)$ and $b=(d-1)/2$ such that
$\varphi$ is $2\pi$-periodic and there is no conical singularity at $\rho=0$.}. The above condition
then just expresses the fact that $\phi_q$ is smooth at $z=1$.
In order to impose this boundary condition, we look for a power series solution to Eq.\ (\ref{AdS_Soliton_Eq}) near 
the IR-floor. We find the Frobenius solution which looks as follows near $z=1$:
\begin{align}
 \phi_{q}^{(1)}(z\rightarrow 1) = 1 + \frac{q^2}{5\alpha^2} (1-z) + \frac{q^4}{100 \alpha^4} (1-z)^2 + \ldots 
    \equiv \sum_{n=0}^\infty c_n (1-z)^n \, ,  \label{SolitonSol3}
\end{align}
where the $c_n$ are also recursively given in appendix \ref{appendix:AdSSoliton}. There, it is also explained that the 
second independent solution near $z=1$ cannot satisfy the above described boundary conditions due
to a divergent term $\sim \log (1-z)$.
Switching from $(z,y)$ to the coordinates $(\rho,\varphi)$  which 
look like usual two-dimensional polar coordinates with origin at $z=1$, 
one can easily verify that the solution (\ref{SolitonSol3}) indeed satisfies the above mentioned 
Neumann condition. 

Knowing that (\ref{SolitonSol3}) is a good solution near $z=1$ and that any solution near $z=0$ can be expressed as a linear combination 
of the solutions (\ref{SolitonSol1}) and (\ref{SolitonSol2}), we need to find out what the global 
solution is. 
We thus write the solution $\phi_q^{(1)}(z)$ satisfying the Neumann condition at 
$z=1$ in the basis of the two fundamental solutions near the boundary:
\begin{align}
 \phi_q^{(1)}(z) = \mathcal{A} \cdot \phi_{q,1}^{(0)}(z) + \mathcal{B} \cdot \phi_{q,2}^{(0)}(z) \label{matching} 
\end{align}
with connection coefficients $\mathcal{A}$, $\mathcal{B}$ which might depend on $q^2/\alpha^2$ but not on $z$. 
Near the boundary, $\phi_{q,1}^{(0)}$ and $\phi_{q,2}^{(0)}$ have the forms (\ref{SolitonSol1}, \ref{SolitonSol2}).
In Eq.\ (\ref{matching}) we kept the normalization $\phi_q^{(1)}(1)=1$. This could be changed arbitrarily, but
as we will see, the Green's function will only depend on the ratio $\mathcal{B}/\mathcal{A}$, so it
would not be affected by another normalization.

We can now use the same prescription as in the case of thermal $\mathcal{N}=4$ SYM 
in order to to calculate the stress-energy correlator $\langle T_2^1 T_2^1 \rangle$ 
\cite{MinkPrescr,SchwingerKeldysch}. For this purpose we need to write down the action of our bulk theory.
For approaches to embed the present theory in string theory, see e.g.\ \cite{Cvetic, Romans, Colgain}. 
We focus on a universal 
gauge/gravity duality and consider just the low energy gravitational sector which is described by the action
\begin{align}
 S = \frac{1}{2\kappa_6^2} \int d^4x\, dy\, dz \; \sqrt{-g} \left( \mathcal{R}- 2\Lambda\right) 
 \label{EinsteinHilbert}
\end{align}
with $\Lambda = -10/\ell^2$ and the six-dimensional gravitational constant $\kappa_6$.
By inserting the metric perturbation given by $\phi$, the part of 
the (on-shell) action which is quadratic in the perturbation can be written as 
\begin{align}
 S_\text{quad.} &= \int \frac{d\omega \, dq}{(2\pi)^2} \, \phi_0(-q) \mathcal{F}(q,z) \phi_0(q) \big{|}_{z=0}^{z=z_0}
   \; + \; \text{contact terms} \notag \\
   \text{with }\mathcal{F} &= -\frac{1}{4\kappa_6^2} \frac{4\pi}{5} \sqrt{-g} g^{zz} \phi_{-q} \partial_z \phi_q 
\end{align}
with the factor $4\pi/5$ coming from integrating out the compact dimension.
The prescription for Lorentzian signature says that we get the retarded Green's
function according to the following rule \cite{MinkPrescr}:
\begin{align}
G^R_{12,12}(\omega,q) &= -2 \mathcal{F}(q,z)\big{|}_{z\rightarrow 0} \notag \\
 &=  \frac{2\pi \ell^4\alpha^4}{\kappa_{6}^2}\frac{\mathcal{B}}{\mathcal{A}} + \text{contact terms} \, ,
\end{align}
where we used the Dirichlet condition $\phi_q(z=\varepsilon \rightarrow 0) = 1$ for the purpose of finding the 
overall normalization. 

The poles of the retarded Green's function are given by the zeros of $\mathcal{A}$. 
On the other hand, setting $\mathcal{A}=0$ in the matching condition (\ref{matching}) would correspond to 
imposing a vanishing Dirichlet condition
at the boundary $z=0$, which defines just the QNM of the AdS soliton geometry.
This conforms with the general arguments in \cite{QNMandHolography}.

If we want to calculate the correlation function in position space, the QNM
become the essential ingredient because we can replace the Fourier integral by a sum over residues.
Since we will be mainly interested in the zero frequency limit, we set $\omega=0$, such that
\begin{align}
\langle \left[ T^1_2(x^3),T^1_2(0)\right]\rangle  &\equiv i G^R_{12,12}(x^3) 
    = \frac{2\pi i\ell^4\alpha^4}{\kappa_{6}^2} \int \frac{dq}{2\pi} \; e^{iqx^3} \frac{\mathcal{B}}{
   \mathcal{A}}(q^2/\alpha^2)\, . \label{greenOO}
\end{align}
Instead of integrating $q$ along the real line, we 
close the contour with a semicircle in the upper complex $q$-plane. 
Due to the Fourier exponential $e^{iqx^3}$, the arc doesn't contribute to the integral and we are left with a sum over QNM residues: 
\begin{align}
 \langle \left[ T^1_2(x^3),T^1_2(0)\right]\rangle &= 
     -\frac{2\pi\ell^4\alpha^4}{\kappa_{6}^2} \sum_{n=1}^\infty e^{-|q_n| x^3} \text{Res}_{q_n}\frac{\mathcal{B}}{\mathcal{A}} 
   \label{AlmostResult}
\end{align}
The calculation of the QN frequencies and of the residues is decribed in the following paragraphs.  
As mentioned above, the simple poles of $1/\mathcal{A}$ are just the QN frequencies. 
Following the general methods in \cite{QNMandHolography},
for these particular values of $q^2/\alpha^2$, the expansion (\ref{SolitonSol3}) of $\phi_q^{(1)}$ is 
normalizable and can be matched smoothly with a linear combination of $\phi_{q,1}^{(0)}$ and $\phi_{q,2}^{(0)}$ over 
the entire interval $z\in (0,1)$; see also \cite{MatchingSol} for an application of similar methods. 
The motivation for this is the observation that the underlying analytic
solution is a power series that converges on the entire interval, 
independent of whether we expand around $z=0$ or $z=1$. 
One can easily check numerically or by investigating the pole structure of the mode equation (\ref{AdS_Soliton_Eq}),
that the radius of convergence of the power series of $\phi_q^{(1)}$ in 
Eq.\ (\ref{SolitonSol3}) reaches $z=0$, such that 
the connection coefficient $\mathcal{A}$ can be found by evaluating the matching equation (\ref{matching}) at $z=0$ with 
the involved functions $\phi_{q,1}^{(0)}$, $\phi_{q,2}^{(0)}$ and $\phi_q^{(1)}$ being given by their
power series expansions:
\begin{align}
 \mathcal{A} = \sum_{n=0}^\infty c_n \, .  \label{A}
\end{align}
The discrete set $\{q_n \; | \; \mathcal{A}(q_n^2/\alpha^2)=0\}$ 
turns out to be purely imaginary. We determine these zeros numerically, using partial sums of the
explicit expansion of $\mathcal{A}$ in Eq.\ (\ref{A}). A WKB estimate for the same eigenvalues has been 
given in \cite{WKBsoliton}, where the set of $m_n^2=-q_n^2$ has been associated with the glueball masses 
in the dual field theory. 
Their result (rewritten in terms of our conventions and parameters) is:
\be
 m_n^2 = -q_n^2 =  n\left(n+\frac{3}{2}\right)  25 \pi \alpha^2 \left( \frac{\Gamma\left(\frac{7}{10}\right)}{\Gamma
    \left(\frac{1}{5}\right)}\right)^2 + \mathcal{O}(n^0) \, . \label{WKBvalues}
\ee

\begin{table}
\centering
\begin{tabular}{|c|c c c c c c|} 
\hline
QNM & $q_1$ & $q_2$ & $q_3$ & $q_4$ & $q_5$ & $q_6$ 
     \\ 
    \hline 
our value$/\alpha$ & $4.061\, i$ & $ 6.688\, i$ & $ 9.249\, i$ & $ 11.786\, i$ & 
    $ 14.313\, i$ & $16.833\, i$ 
    \\             
WKB estimate$/\alpha$ & $ 3.96\,  i$ & $6.63\,  i$ & $ 9.21\,  i$ & $11.75\, i$ & 
    $14.29\, i$ & $16.81\, i$ 
     \\
    $\text{Res}_{q_n} \; \mathcal{B}/\mathcal{A}$ & $-21.08$ & $-244.4$ & $-1219$ 
     & $-4072$ & $-1.07 \cdot 10^{4}$ & $-2.41 \cdot 10^{4}$  
     \\ 
\hline
\end{tabular}
\caption{The first two lines show the values of the lowest QN frequencies in the six-dimensional AdS-soliton, using our
matching method and the WKB estimate from \cite{WKBsoliton}, respectively. 
The third line shows the residues of $\mathcal{B}/\mathcal{A}$ at these points.}
\label{tab:QNM}
\end{table}
The first QN frequencies are listed in table \ref{tab:QNM}.
Even beyond the shown accuracy, the values that we obtain from our matching method agree  
precisely with what we find by just using a finite differences algorithm to solve 
Eq.\ (\ref{AdS_Soliton_Eq}) numerically.
We observe that the WKB results agree with these exact values to an accuracy which is in accordance with 
Eq.\ (\ref{WKBvalues}), becoming better for larger $n$.
Note that the complex conjugates of all the $q_n$ are also zeros of $\mathcal{A}$. However, we will not 
need them because we close the contour of the integral in Eq.\ (\ref{greenOO}) in the upper half plane.

\begin{figure}
 \begin{center}
  \label{fig:ads_soliton}
    \includegraphics[width=.7\textwidth]{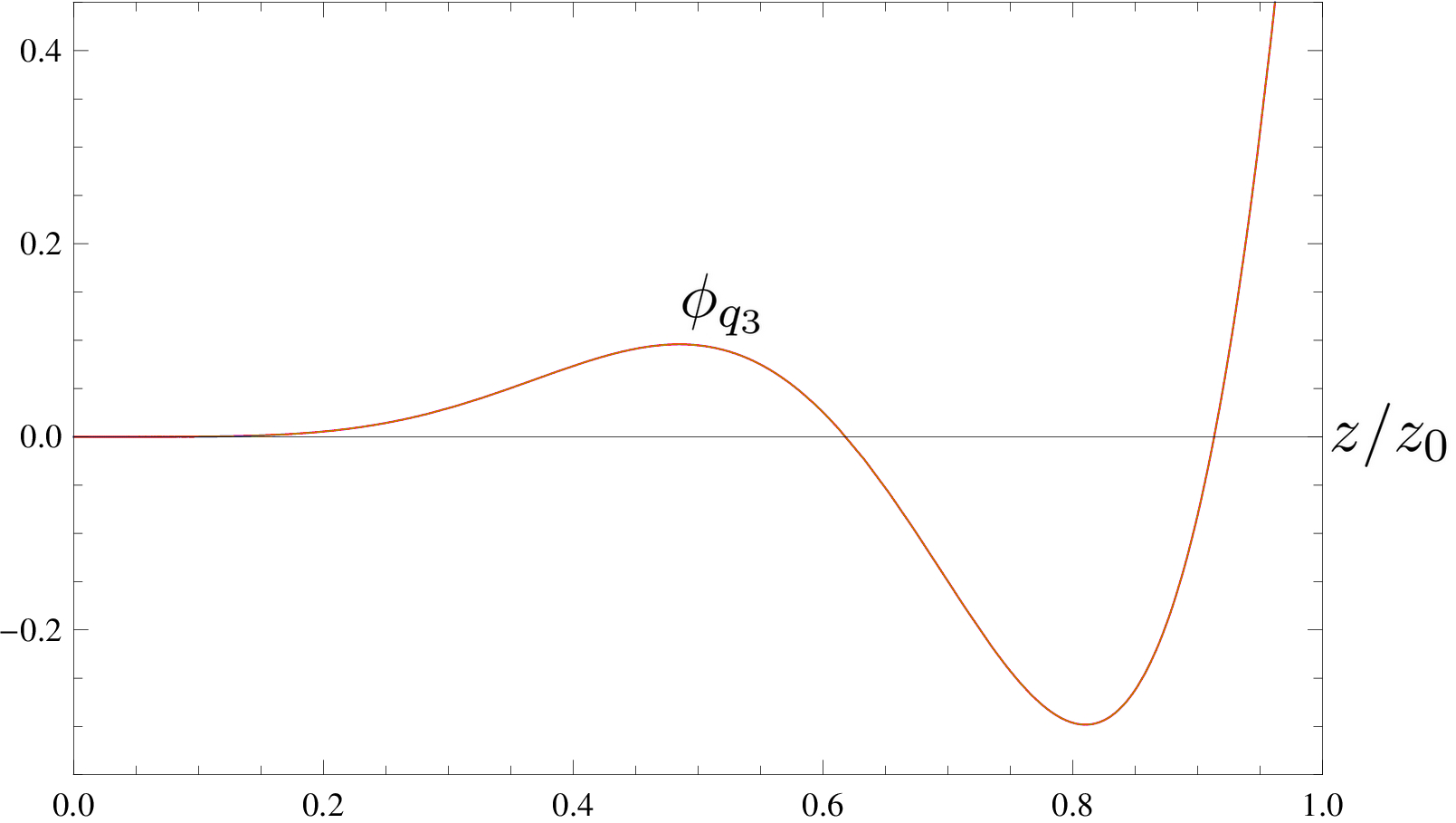}\\ \vspace{.4cm}
    \includegraphics[width=.7\textwidth]{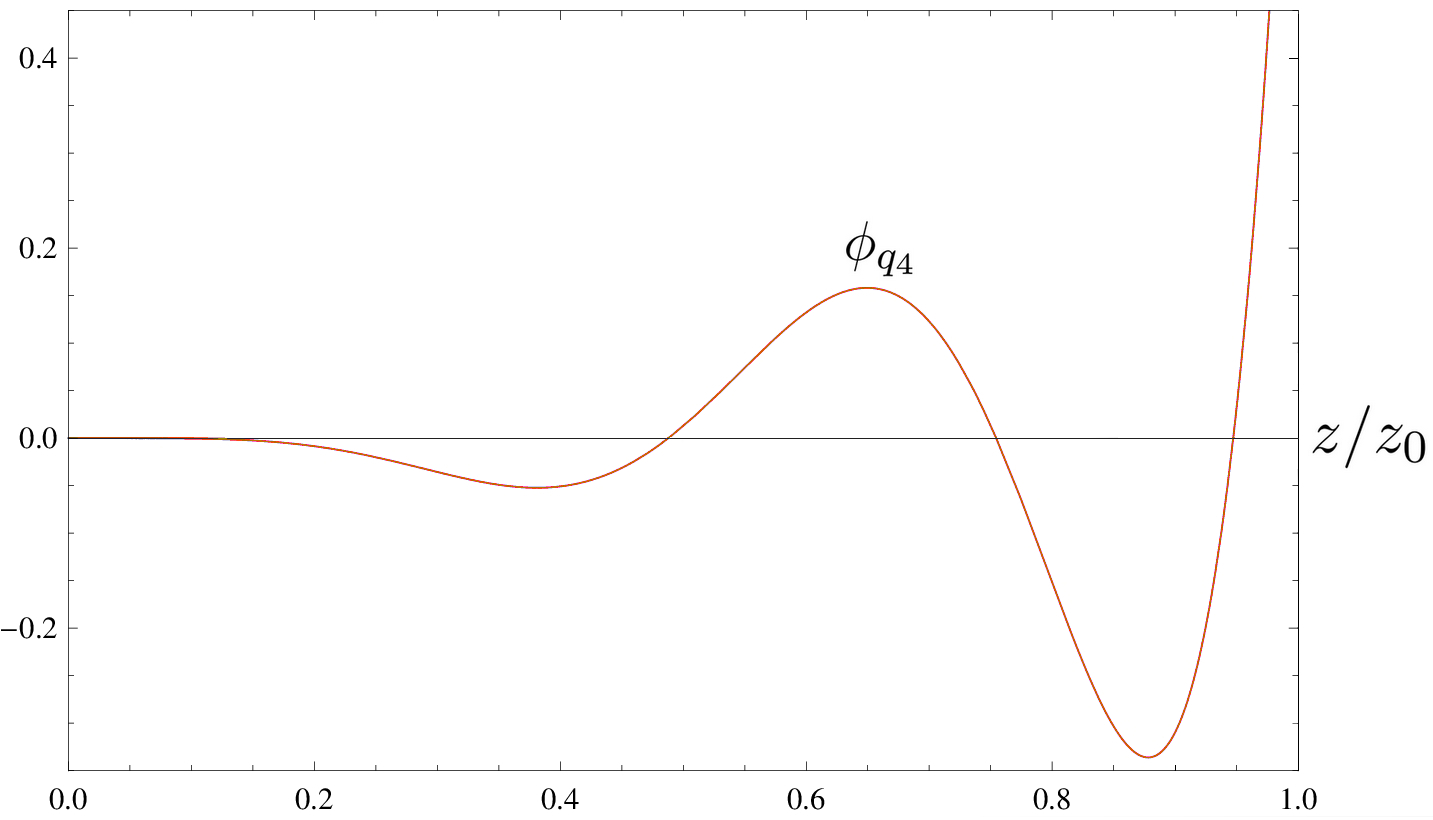} 
    \caption{The quasinormal mode functions of the six-dimensional 
     AdS soliton, $\phi_q$, plotted for the eigenvalues $q_3$ and $q_4$. The overall
     scaling has been fixed by normalizing $\phi_{q_n} (z=1) = 1$ as in Eq.\ (\ref{matching}). 
     For the discrete values $q\in \{q_n\}$ the expansion 
     around $z=0$ and the one around $z=z_0$ match over the entire interval $[0,1)$. For all other values of $q$ 
     such a matching is
     not compatible with the normalizability condition $\phi_q(0)=0$. 
}
 \end{center}
\end{figure}

The functions $\phi_{q_n}(z/z_0)$ for $n=3,5$ are plotted in fig.\ \ref{fig:ads_soliton}.
The expansion around $z=0$ is shown on the interval $[0,0.95]$, and the expansion around $z=1$ is shown 
on $[0,1]$. We cannot distinguish them in the plots since they
match perfectly over the entire common interval when $q\in\{q_n,q_n^*\}_n$. 
For all other values of $q$ the boundary condition $\phi_q(0)=0$ makes it impossible to match the two expansions.

We can also express $\mathcal{B}$ as a function of $\mathcal{A}$. 
Since all the power series expansions converge at $z=1/2$, we can evaluate Eq.\ (\ref{matching}) at this point\footnote{Other 
points in $(0,1)$ lead to the same result, of course. The convergence might be worse, however.}, and find
\begin{align}
 \mathcal{B} = \frac{\left(\sum_{n=0}^\infty c_n 2^{-n} \right) - \mathcal{A} \cdot \left( \sum_{n=0}^{\infty}
       a_n 2^{-n} \right)}{\left(\sum_{n=5}^\infty b_n 2^{-n}\right)} \, . \label{BofA}
\end{align}

We can now calculate the residues of $\mathcal{B}/\mathcal{A}$
at the QNM poles. We take Eq.\ (\ref{A}) and plot $(q-q_n)/\mathcal{A}$, 
which is a smooth function in the vicinity of $q_n$, 
and determine the value
at $q_n$ with high numerical precision. This result is multiplied with the value of $\mathcal{B}$ at the particular
point, $\mathcal{B}(q_n^2/\alpha^2)$. This latter value can easily be obtained from Eq.\ (\ref{BofA}) with the second summand 
in the numerator, which is proportional to $\mathcal{A}$, set to zero. 
This method works very well at least for the lower lying QN frequencies.
The values of the first six residues are shown in the last line of table \ref{tab:QNM}. 
Using these results, we can evaluate the expression in Eq.\ (\ref{AlmostResult}):
\begin{align} 
\langle \left[ T^1_2(x^3),T^1_2(0)\right]\rangle=-\frac{2\pi\ell^4\alpha^4}{\kappa_{6}^2} 
\left[ e^{-4.061 \alpha \, x^3} \cdot (-21.08) + \ldots \right]\, . \label{Result_ads_soliton}
\end{align}
This result confirms our expectation of an exponentially decaying correlation function 
in the long wavelength limit. Physically this means that shear diffusion to infinity is 
strongly supressed. 
The contribution of the lowest QNM dominates the sum over exponentially decaying terms. Although the values of the residues
grow (see table \ref{tab:QNM}), the exponentials make every higher QN frequency $q_n$ completely insignificant compared to the 
contribution of $q_{n-1}$.

\section{Instability of the Schwarzschild Soliton String}
\label{sec:Instability}
We will now introduce a modification of the AdS soliton which contains a black hole 
in the boundary metric.
Before we calculate Green's functions and holographic transport properties in this novel geometry, it will turn out to be
useful to carry out a stability analysis in terms of linearized perturbations. 

\subsection{Generalizations of the AdS Soliton}
It has been noted in \cite{MarolfAdSString} that the $d$-dimensional AdS soliton metric (\ref{AdS_Soliton}) can very easily be
generalized. In fact, one can replace the Minkowski metric $\eta_{\mu\nu}$ in (\ref{AdS_Soliton}) by any 
Ricci flat metric $g_{\mu\nu}(x^\mu)$ and the resulting metric is still a solution of Einstein's vacuum equations: 
\be
ds^2 = \frac{r^2}{\ell^2} \left[ g_{\mu\nu}dx^\mu dx^\nu + \left( 1- \frac{r_0^{d-1}}{r^{d-1}} \right) d\tau^2 \right] 
   + \left( 1- \frac{r_0^{d-1}}{r^{d-1}} \right)^{-1} \frac{\ell^2}{r^2} dr^2\, . \label{SchwSolitonStringMetricOrig}
\ee
Again, $\tau\sim \tau + 4\pi \ell^2/(d-1)r_0$ needs to be periodically identified.

If we choose for $g_{\mu\nu}$ the four-dimensional Schwarzschild metric, Eq.\ (\ref{SchwSolitonStringMetricOrig})
describes the Schwarzschild soliton string. The geometry looks like Schw${}_4\times S^1$ stretched out in a 
string along a AdS radial direction $r$ that caps off smoothly at a finite value $r=r_0$.
The two relevant physical scales are the Schwarzschild radius $r_\text{s}$ of the black hole and the 
radius of the compact dimension, i.e.\ $L_\tau = 2\ell^2/5r_0$.
In order to simplify calculations considerably, we perform the following 
transformations:
\be
  r \longrightarrow \frac{r_0}{z}\, , \quad \tau\longrightarrow \frac{\ell^2}{r_0} y\, , \quad t \rightarrow r_\text{s} \bar t,\,
   \quad \rho \longrightarrow  r_\text{s} \bar r\, , \label{changeCoords}
\ee
where $r$ is the original radial AdS coordinate, and $\rho$ is the original radial coordinate in Schw${}_4$. 
This brings the Schwarzschild soliton string metric in the form
\begin{align}
ds^2 = \frac{\ell^2}{z^2} &\left\{ \alpha^2 \left[ -\left(1-\frac{1}{\bar r}\right) d\bar t^2 + \left(1-\frac{1}{\bar r}\right)^{-1}
    d\bar r^2 + \bar r^2 d\Omega^2 \right] + (1-z^5) dy^2 + \frac{dz^2}{(1-z^5)} \right\} , \label{SchwSolStringFinal}
\end{align}
where we defined the parameter $\alpha \equiv r_0 r_\text{s}/\ell^2$. 

The Schwarzschild soliton string is supposed to describe the bulk dual of a strongly coupled
field theory in a Schwarzschild black hole background. 
It inherits the important property of the AdS soliton that it caps off smoothly at the IR floor deep in AdS. 
A more naive choice for a gravity dual of a field theory in a black hole background would be the 
AdS black string \cite{GregLafl:AdSString}. 
However, a serious problem would be that the AdS black string is nakedly singular at the end point along the string direction.
Also, as discussed in \cite{islam},
trying to cover this singularity by a horizon does not eliminate the presence of nakedly singular surfaces.
The Schwarzschild soliton string clearly solves these problems: There is no naked singularity due to the
special geometry inherited from the AdS soliton. 


\subsection{Reduction to Gregory-Laflamme Instability}
We want to show that a decomposition of a perturbation 
of the Schwarzschild soliton string geometry in tensor, vector and scalar modes with respect
to the base manifold $\mathfrak{B} =\text{Schw}_4$ produces an instability in the tensor sector. The 
decomposition of the metric (\ref{SchwSolStringFinal}) reads
\be
ds^2 = g_{AB} dx^A dx^B + a^2(x^A) ds_\text{Schw.}^2 \; , \qquad ds_\text{Schw.}^2 = \hat g_{\mu\nu}dx^\mu dx^\nu \, , \label{metricgeneral}
\ee
where $\mu,\nu$ run over the indices of the four-dimensional Schwarzschild metric $\hat g_{\mu\nu}$ 
in the coordinates of (\ref{SchwSolStringFinal}),
$g_{AB}$ describes the two-dimensional
orbit space which is parameterized by $(y,z)$, and $a(z) \equiv \ell \alpha/z$.
Let us start with a transverse tracefree (TTF) tensor perturbation of the form 
\be
g_{ab} \longrightarrow g_{ab} + h_{ab}, \qquad h_{ab} = \bp h_{\mu\nu} & 0 \\ 0 & 0 \ep \, , 
 \qquad h^a{}_a =0 = h^{ab}{}_{,b} \, ,
\ee
where $a,b=0,\ldots,5$ run over all coordinates. For a solution of the Einstein equations of the form
\be
ds^2 = a^2(z) ds_\text{Schw.}^2  + \xi(z) dy^2 + \eta(z) dz^2 \, ,
\ee
the linearized Einstein equations read 
\begin{align}
0 &= \underbrace{\left( -\delta^c_a \delta^d_b \Box - 2 R_a{}^c{}_b{}^d \right) h_{cd}}_{\equiv \Delta_L h_{ab}} + 2R^c{}_{(a}h_{b)c} 
    +2 \nabla_{(a}\nabla^c h_{b)c} -\nabla_a \nabla_b h  \notag \\
  &= \frac{1}{a^2} \hat \Delta_L h_{\mu\nu} - \frac{1}{\eta} \partial_z^2 h_{\mu\nu} + \frac{1}{2\eta} \left[ \frac{\eta'}{\eta}
   - \frac{\xi'}{\xi} \right] \partial_z h_{\mu\nu} \notag \\
    &\quad  + \frac{2}{\eta} \left[ \frac{a''}{a} + \left( \frac{a'}{a} \right)^2
   + \frac{1}{2} \frac{a'}{a} \left( \frac{\xi'}{\xi} - \frac{\eta'}{\eta} \right) \right] h_{\mu\nu}    
   - \frac{1}{\xi} \partial_y^2 h_{\mu\nu}\, ,\label{linEFE2}
\end{align}
where $\Delta_L$ is the Lichnerowicz Laplacian and $\hat \Delta_L$ is the Lichnerowicz operator on the 
four-dimensional base manifold $\mathfrak{B}$.
We choose a harmonic dependence on $y$.
Furthermore, the $z$-dependence is the same for each component $h_{ab}$.
Writing $h_{\mu\nu}(x^\mu,y,z) = \chi_{\mu\nu}(x^\mu) e^{i\nu y} H(z)/z^2$, we find that
the $z$-dependence of this equation separates:
\begin{align}
 0 &= \left( \hat\Delta_L + m^2 \right) \chi_{\mu\nu} \, , \label{eq1}\\
 0 &= H''(z) - \left( \frac{5z^4}{1-z^5}+\frac{4}{z} \right) H'(z) + 
     \left(\frac{m^2}{\alpha^2} \frac{1}{1-z^5} - \frac{\nu^2}{(1-z^5)^2}\right)H(z)\, ,  \label{eq2}
\end{align}
where $m^2$ is the constant that comes from separating the variable $z$.
For $\nu=0$ (no excitation in the compact dimension) 
the second of these equations is exactly the same as that of a scalar propagating in the AdS soliton, Eq.\ (\ref{AdS_Soliton_Eq}).
We can therefore use the results that we derived in section \ref{sec:ads_soliton_calc}: There is an infinite tower
of discrete values of $m^2/\alpha^2>0$ for which Eq.\ (\ref{eq2}) has a regular solution. The existence of an unstable mode
thus depends on the existence of such a mode which solves Eq.\ (\ref{eq1}). But Eq.\ (\ref{eq1}) is just the 
equation that governs perturbations of the five-dimensional Schwarzschild black string, i.e.\ the well-known
Gregory-Laflamme problem \cite{GL1}. We assume that the time dependence of $\chi_{\mu\nu}$ is of the form 
$e^{\Omega t}$. 
Using the same numerical methods as for the solution of Eq.\ (\ref{eq2}), we can show that 
there exists a threshold mode with $\Omega = 0$ which
corresponds to the maximum boundary mass $m_\text{max}^2\approx 0.768$ such that all 
$0<m^2<m_\text{max.}^2$ give modes which grow in time. This is in accordance with the results of \cite{GL1}.

However, we want to investigate Eq.\ (\ref{eq1}) in some more detail since we will need the form of the analytic solution 
in section \ref{sec:StringCorrelators}.
We use the spherically symmetric threshold Gregory-Laflamme ansatz:
\be
(\chi_{\mu\nu}) = \bp h_0(x) & h_1(x) & 0 &0\\ h_1(x) & h_2(x) & 0& 0 \\ 0&0&K(x)&0
   \\ 0&0&0& K(x)\; \text{sin}^2\theta \ep \, ,
\ee
where we introduced the coordinate $x\equiv 1/\bar r$ with $0<x\leq 1$. 
Imposing the TTF gauge condition on this ansatz, the perturbation equations (\ref{eq1}) reduce
to the following set of equations which form an effectively one-dimensional problem:
\begin{align}
 &x^4 \left(3 x^2-5 x+2\right) h_2'' - 2x^3  \left(3 x^2-6 x+2\right) h_2' -\left(8 x^3+m^2 (2-3 x)\right) h_2= 0\, , \label{GLrev1}\\
 &\qquad h_0(x) = \frac{1}{3 x-2}\left(2 x(x-1)  h_2'-(5x-6)h_2\right)\, , \label{relh0h2} \\
 &\qquad K(x) = \frac{1}{3 x-2}\left(-x(x-1)  h_2'+(x-2)h_2 \right)\, , \\
 &\qquad h_1(x) = 0\, .\label{GLrev4}
\end{align}
Solving the first of these equations (e.g.\ numerically with a finite differences algorithm and 
regular boundary conditions) indeed yields as the
only non-negative eigenvalues of this system the zero mode $m^2=0$ and the Gregory-Laflamme 
threshold mode $m^2 \approx 0.768$.

\subsection{Relation Between Physical Scales}
\begin{figure}
 \begin{center}
  \label{fig:instability}
    \includegraphics[width=.95\textwidth]{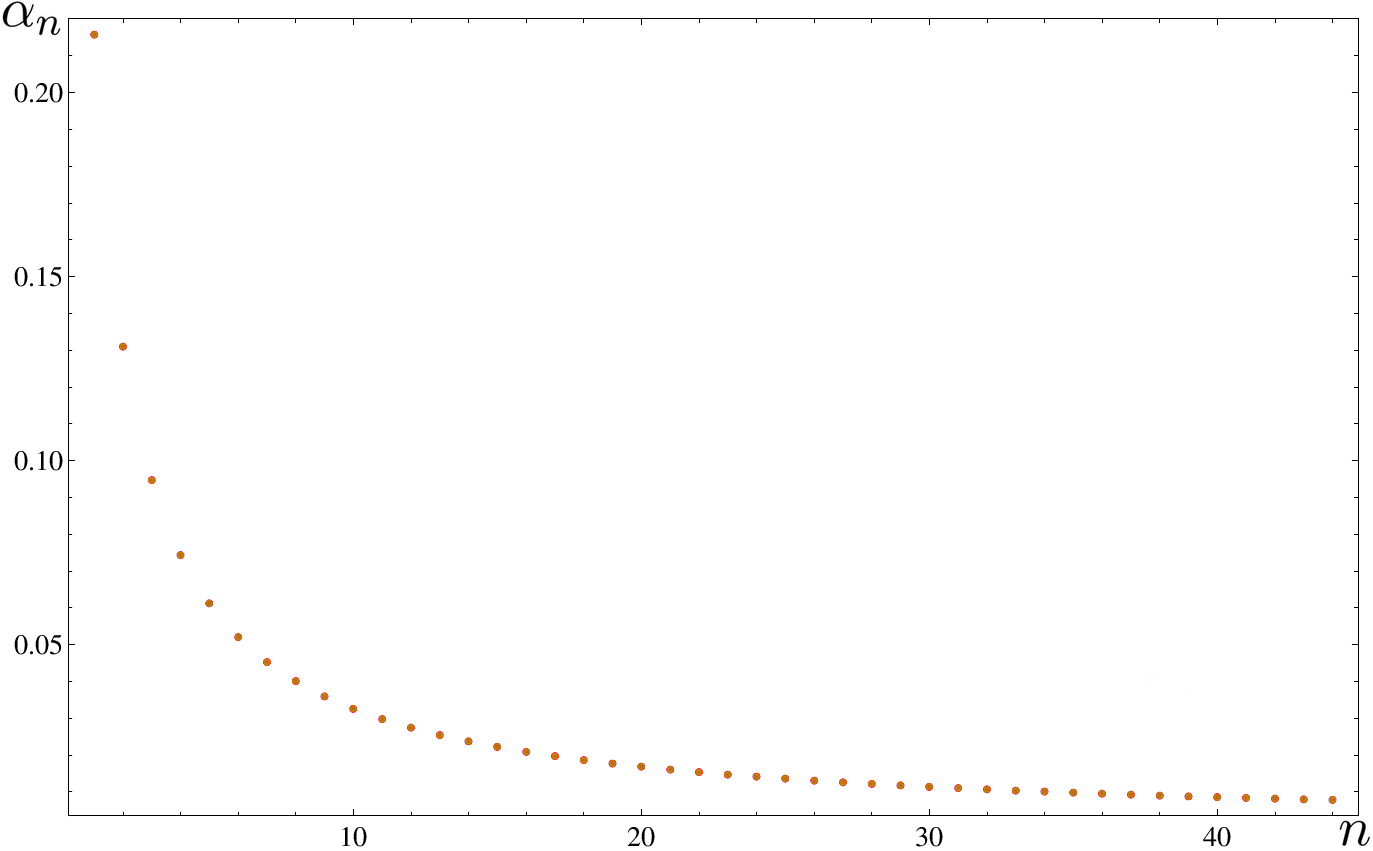} 
    \caption{The values $\alpha\equiv r_0 r_s /\ell^2$ for which the perturbation 
    equations (\ref{eq1}, \ref{eq2}) have a solution. The leftmost point corresponds to the critical 
    value of $\alpha$ for which an instability sets in.     
}
 \end{center}
\end{figure}

We can now draw conclusions about the critical values of the size of the compact dimension and of the Schwarzschild radius of the 
boundary black hole. 
To this end, we need to find the simultaneous eigenvalues $m^2$ of Eq.\ (\ref{eq1}) and $m^2/\alpha^2$
of Eq.\ (\ref{eq2}). 
Therefore, we need to combine the threshold eigenvalue $m_\text{max.}^2 \approx 0.768$ of the Lichnerowicz operator on the 
background metric with the discrete values for
$m^2/\alpha^2 \equiv -q^2/\alpha^2$ from section \ref{sec:ads_soliton_calc}. This yields
\be
\alpha \equiv \frac{r_0 r_\text{s}}{\ell^2} 
  =\sqrt{\frac{m_\text{max.}^2}{(-q^2/\alpha^2)}} \in \left\{ 0.216, \; 0.131, \; 0.095, \; 0.073,\ldots \right\}\, .
\ee
This list continues and, in fact, it gives an infinite tower of discrete values for $\alpha$ which asymptotically approach $0$. See fig.\ \ref{fig:instability}

The largest value of $\alpha$ corresponds to the most unstable mode. Indeed, since we used the critical value $m_\text{max.}$ for
which an instability can occur, the values of $\alpha$ for unstable linearized tensor modes cannot 
exceed the value $\alpha_\text{crit.}\approx 0.216$.
Writing this result in terms of the physically relevant scales $r_\text{s}$ and the radius of the compact dimension $L_\tau = 
2\ell^2/5r_0$, we conclude that the critical ratio for an instability to occur is 
\be
 \left(\frac{r_\text{s}}{L_\tau}\right)_\text{crit.} = 
    \frac{5}{2}\cdot\left(\frac{r_0 r_\text{s}}{\ell^2}\right)_\text{crit.}  \approx 0.539 \sim \mathcal{O}(1)\, .
\ee 
It is the interplay between these two parameters which determines the stability of the Schwarzschild soliton string.
If the horizon radius $r_\text{s}$ is small compared to the size of the compact circle
(i.e.\ not bigger than allowed by the above equation), 
then the Schwarzschild soliton string is unstable. Note that we cannot straightforwardly reverse this reasoning
without studying the vector and scalar sectors of perturbations. 
In appendix \ref{appendix:vectorscalar} we
carry out the analyses of linearized perturbations in the vector and scalar sectors of the decomposition 
of the Schwarzschild soliton string with respect to $\mathfrak{B}=\text{Schw}_4$. 
We demonstrate that neither vector nor scalar modes give rise to an instability. 

Holographically, we interpret the instability in terms of the confinement scale of the gauge theory. 
Since we associate the temperature of the field theory with the inverse period of the compact direction $\tau$,
we obtain a temperature for the deconfinement transition
$T_\text{dec.}=(1/L_\tau)_\text{crit.}$ for given $r_\text{s}$. This confirms the conjecture of \cite{MarolfAdSString}
in a quantitative way.

Note that if we had not chosen $\nu=0$ to solve Eq.\ (\ref{eq2}), the QN frequencies would have been all shifted to 
a bigger value such that the critical value of $\alpha$ for an instability to occur would have turned out to be smaller. 
This can easily be confirmed by solving Eq.\ (\ref{eq2}) numerically for different values of $\nu$. 
Therefore, in order to find the true critical (i.e. \ maximal) value of $\alpha$ it was safe to assume that the perturbation 
is not excited along the compact dimension.

\section{Stress Tensor Correlators from the Schwarzschild Soliton String}
\label{sec:StringCorrelators}
We turn now to the task of calculating stress-energy two-point functions in the boundary field theory dual to the
Schwarzschild soliton string. The analysis is complicated due to a lack of translational 
invariance of the boundary field theory metric. First of all,
we need again to decide how we use the symmetries of the problem to decompose the perturbation. 
As a starting point, we take the Gregory-Laflamme mode and Eqs.\ (\ref{GLrev1}-\ref{GLrev4}) from the previous section. 
This corresponds to a TTF tensor mode with respect to the base manifold $\mathfrak{B}=\text{Schw}_4$, and it will 
eventually allow us to calculate the correlators
\be
\langle T_{tt} T_{tt} \rangle \, ,\quad\langle T_{\bar r \bar r} T_{\bar r \bar r} \rangle \, ,\quad 
\langle T_{\theta\theta} T_{\theta\theta} \rangle \, ,\quad \langle T_{\phi\phi} T_{\phi\phi} \rangle \,  ,
\ee
the first of which is particularly interesting: Since $T_{tt}$ is the energy density, the associated two-point function
describes heat transport in the field theory.
According to the dictionary of gauge/gravity duality, we need to find the solution to the equation of motion that
separated into the equations (\ref{eq1}) (or equivalently Eq.\ (\ref{GLrev1})) and (\ref{eq2}).
Then we need to calculate the on-shell action quadratic in the gravitational perturbation, 
and take appropriate functional derivatives. 

To our knowledge, an analytic solution to Eq.\ (\ref{GLrev1}) does not exist. However, two observations suffice to 
calculate the desired Green's functions in a suitable limit. First, we write the equation in Sturm-Liouville form: 
\begin{align}
 \partial _x\left(\frac{(1-x)^2}{(2-3 x)^2 x^2}h_2'\right) - \left(\frac{8(1-x)}{x^3 (2-3 x)^3}
    + \frac{(1-x)}{x^6 (2-3 x)^2}  m^2 \right)h_2  = 0  \, . \label{SLform}
\end{align}
Then, Sturm-Liouville theory 
ensures that whatever the exact solution to this equation is,
in an appropriate normalization the solutions to different eigenvalues $m^2$ are orthonormal with respect
to the inner product 
\be
\int dx \; w(x) h_2^{(m_1)} h_2^{(m_2)}  = \delta_{m_1^2 m_2^2} \, ,
\ee
where the weight function $w(x)$ is given by the factor multiplying the eigenvalue in Eq.\ (\ref{SLform}):
\be
 w(x) = \frac{(1-x)}{x^6 (2-3 x)^2} \, .\label{weight}
\ee
This is the first 
observation. The second observation concerns the \textit{asymptotic} solution to Eq.\ (\ref{GLrev1}). For the 
solution which is normalizable as $x\rightarrow 0$ and regular near the
horizon, we find the following asymptotics: 
\be
 h_2^{(m)}(x) \sim \left\{ \begin{aligned}
                             &C_0 \, e^{-|m|/x}\, x^{2+|m|/2} \left[1 + \mathcal{O}\left(x\right) \right]
                                         & \text{ for } x \rightarrow 0 \, ,\\
                             &C_1 \, \left[ 1 - \tfrac{1}{2} \left(8-m^2 \right) (1-x) + \mathcal{O}\left( (1-x)^2 \right)
                                         \right]            & \text{ for } x\rightarrow 1 \, ,
                          \end{aligned}
   \right. \label{asymptLsgSL} \,
\ee
We can recover the full tensor mode 
$h_{\mu\nu}(x^\mu,y,z)=\chi_{\mu\nu}(x,\theta)e^{i\nu y} H(z)/z^2$ by integrating over the boundary masses. In particular, the 
time-independent and $y$-homogeneous $11$-component is given by
\be
h_{11}(x,z) =  \int d(m^2) \; c^{(m)} \,\frac{H^{(m)}(z)}{z^2}\,h_2^{(m)}(x) \, , \label{expan}
\ee
where $H^{(m)}(z)$ is the function $H(z)$ for a particular eigenvalue $m^2$. Note that the
correlators which we will calculate refer to the Schwarzschild soliton string in its stable phase. 
Therefore the integral (\ref{expan}) runs over $m^2 > m_\text{max.}^2$, where $m_\text{max.}$ is the threshold
mode for the instability to occur. 
Eq.\ (\ref{expan}) is the analog of the spatial Fourier transform that we used in the translationally invariant case (section 
\ref{sec:ads_soliton_calc}):
The functions $h_2^{(m)}$ take the role of the orthonormal set of base functions, and the $c^{(m)}$ 
correspond to the boundary values with respect to which we will take functional derivatives. We define them in analogy to 
what we did in the case of the AdS soliton, such that the normalization of the ``mode'' function $H^{(m)}(z)$ is given 
by $H^{(m)}(0)=1$. Let us again assume $\nu=0$, i.e.\ no excitation in the compact dimension. 
Then the Eq.\ (\ref{eq2}) for $H^{(m)}(z)$ is  
the same equation as for the mode $\phi_q(z)$ in the case of the AdS soliton. Thus we conclude
from our analysis in section \ref{sec:ads_soliton_calc} that the properly normalized solution is
\begin{align}
H^{(m)}(z) &=  \left( 1 + \frac{m^2}{6\alpha^2} z^2 + \ldots \right) + \frac{\mathcal{B}}{\mathcal{A}}(-m^2/\alpha^2)
  \left(z^5 - \frac{m^2}{14\alpha^2} z^7 + \ldots \right)\, .
\end{align}

These information can be used to determine the Green's function that is induced by $h_2$. In principle, we proceed
as in the case of thermal $\mathcal{N}=4$ SYM and the AdS soliton (c.f.\ \cite{Viscosity} and section \ref{sec:ads_soliton_calc},
respectively). 
However, the details are 
more complicated due to a lack of translational invariance. Using the equations of motion,
the part of the six-dimensional Einstein-Hilbert action which is quadratic in $h_2$ 
can again be written as a five-dimensional integral over the 
AdS boundary term.  We start by identifying the relevant terms in the quadratic part of the action (\ref{EinsteinHilbert}):
\begin{align}
 S_\text{quad.}&= \frac{1}{2\kappa_{6}^2} \int dt\,dx\,d\theta \,d\phi\, dy\,dz \; 
    2\ell^4\alpha^4 \sin\theta \bigg{[} \frac{1-z^5}{z^4} \left( 3H'(z)^2   +4H(z)H''(z) + \ldots \right)\bigg{]} \notag \\
   &\qquad\qquad\quad \times \bigg{[} \mathcal{C}_1(x)h_2(x)^2  + \mathcal{C}_2(x) h_2(x)h'_2(x) 
              + \mathcal{C}_3(x)h_2'(x)^2\bigg{]} \, ,  \label{quadA}
\end{align}
where the dots stand for terms which do not contain the right number of derivatives in $z$, so they will be
irrelevant in the eventual application of the gauge/gravity recipe. Furthermore, we have defined
\begin{align}
 \mathcal{C}_1(x) \equiv \frac{2\left(9 x^2-20x +12\right)}{x^4(2-3x)^2} \;,\quad
 \mathcal{C}_2(x) \equiv -\frac{4 x\left(3 x^2-7x+4\right)}{x^4(2-3x)^2} \;,\quad 
 \mathcal{C}_3(x) \equiv \frac{3x^2(x-1)^2}{x^4(2-3x)^2} \, .
\end{align} 
We proceed by using partial integration in order to remove all derivatives of $h_2$ and eventually make the $x$-integration
trivial. 
The first step consits of integrating by parts the term $\sim (h_2')^2$. This yields a new term which contains a second derivative, 
$h_2''$. If we express the full solution $h_2(x)$ as an integral over ``mode functions'' 
$h_2^{(m)}$ as in Eq.\ (\ref{expan}), 
the resulting integral reads
\begin{align}
 S_\text{quad.}&= \frac{16\pi^2\ell^4\alpha^4}{5\kappa_{6}^2} \int dx\, dz \, d(m_1^2) \, d(m_2^2) \; c^{(m_1)} c^{(m_2)} \notag \\
   &\qquad\qquad\qquad\;\; \times  \bigg{[} \frac{1-z^5}{z^4} \left( 3H^{(m_1)}H^{(m_2)}   +4H^{(m_1)}H^{(m_2)}{}'' + \ldots \right)\bigg{]} \notag \\
   &\qquad\qquad\qquad\;\; \times \bigg{[} \mathcal{C}_1 h^{(m_1)}_2h^{(m_2)}_2
         + \left(\mathcal{C}_2- \mathcal{C}_3' \right) h^{(m_1)}_2h^{(m_2)}_2{}' 
         - \mathcal{C}_3 \, h_2^{(m_1)} h_2^{(m_2)}{}'' \bigg{]} \, ,  
\end{align}
where also the trivial integrals over $\theta$, $\phi$, $y$ have been performed, and the time-integral has
been deleted. The time integral 
can be omitted because we consider time-independent solutions, i.e.\ the final Green's function will not be 
localized in time.

If we replace $h^{(m_2)}_2{}''$, using the equation of motion (\ref{GLrev1}), there are only terms left which look like
either $\sim h^{(m_1)}_2h^{(m_2)}_2$ or $\sim h^{(m_1)}_2h^{(m_2)}_2{}'$. 
Since the coefficient of the $h^{(m_1)}_2h^{(m_2)}_2{}'$-term depends on $m_1$ and $m_2$ in the same way 
as in the definition (\ref{expan}), we can absorb the integrals
over the boundary masses such that this term becomes $\sim h_2 h_2'=\partial_x [(h_2)^2/2]$ and another partial integration 
in $x$ can be performed. This finally gives an expression that does not contain any derivatives of $h^{(m_i)}_2$:
\begin{align}
S_\text{quad.}^\text{(on-shell)}&= \frac{16\pi^2 \ell^4\alpha^4}{5\kappa_{6}^2} \int dx\,dz\, d(m_1^2) \, d(m_2^2) \; c^{(m_1)} c^{(m_2)}\,
      \left[ -3 m_2^2 \, w(x) \,h^{(m_1)}_2(x) h^{(m_2)}_2(x)\right] \notag \\
     & \qquad\qquad\qquad\;\;\;\times 
    \bigg{[} \frac{1-z^5}{z^4} \left( 3H^{(m_1)}{}'H^{(m_2)}{}'   +4H^{(m_1)}H^{(m_2)}{}'' + \ldots \right)\bigg{]}  \, .
\end{align}
 The function $w(x)$ is precisely the weight function from Eq.\ (\ref{weight}) for which
Sturm-Liouville solutions $h_2^{(m)}(x)$ are orthonormal. The integral over $x$ thus yields a $\delta$-function in 
the eigenvalue, $\delta(m_1^2-m_2^2)$. 

We still have to integrate out the $z$-dependence, and find the boundary value of the action. First, we integrate by parts
the term $\sim H^{(m)}H^{(m)}{}''$, such that it contributes with a negative sign to the part $\sim(H^{(m)}{}')^2$. This produces 
two more terms, one of which is a boundary term that is precisely cancelled by the Gibbons-Hawking contribution to the 
Einstein-Hilbert action (\ref{EinsteinHilbert}). 
The second unwanted term is $\propto H^{(m)}H^{(m)}{}'$ and it is therefore not relevant for the purpose of applying the gauge/gravity recipe. 
The remaining term $\sim (H^{(m)}{}')^2$ can be integrated as in the translationally invariant case, and we find 
\begin{align}
S_\text{quad.}^\text{(on-shell)}&=  \frac{48\pi^2 \ell^4\alpha^4}{5\kappa_{6}^2} 
     \int  d(m^2) \; (c^{(m)})^2\, m^2 \, \frac{1-z^5}{z^4} H^{(m)}(z) H^{(m)}{}'(z) \bigg{|}_{z=0}^{z=1} + \text{contact terms}\, .
\end{align}
Following the gauge/gravity recipe, we take functional derivatives and obtain the final result for the 
Green's function in position space\footnote{Since we work 
in Lorentzian signature, we would actually have to follow the recipe \cite{MinkPrescr} and throw away the contribution
at $z=1$. Because this contribution is zero anyway, the result would be the same.}:
\begin{align}
 G^R_{11,11}(x_1,x_2) &= - \int d(m_1^2) \, d(m_2^2) \; \frac{\delta^2 S_\text{quad.}^{\text{(on-shell)}}}{\delta c^{(m_1)}
     \delta c^{(m_2)}}\cdot h_2^{(m_1)}(x_1) h_2^{(m_2)}(x_2)  \notag \\
 &= -\frac{96\pi^2 \ell^4\alpha^4}{\kappa_{6}^2}   \int d(m^2) \; m^2 \,
         h_2^{(m)}(x_1) h_2^{(m)}(x_2) \, \frac{\mathcal{B}}{\mathcal{A}}(-m^2/\alpha^2) \notag \\
 &= -\frac{192i\pi^3 \ell^4\alpha^4}{\kappa_{6}^2}  \sum_{n=1}^\infty  m_n^2 \, h_2^{(m_n)}(x_1) h_2^{(m_n)}(x_2)
        \,\text{Res}_{m_n} \left( \frac{\mathcal{B}}{\mathcal{A}} (-m^2/\alpha^2) \right) \, ,
         \label{eq413}
\end{align}
where we evaluated the integral over $m^2$ as follows: The poles of $\mathcal{B}/\mathcal{A}$ lie on the integration contour. 
Therefore, the integral is given by its principal value which can be evaluated by deforming the contour such that it
goes around the poles in small semicircles in the lower half plane. Closing the contour at infinity, we can apply
the Cauchy residue theorem similar to what we did in the case of the AdS soliton. 
In order to determine the asymptotic behavior of the bi-tensor $G^R_{ab,cd}(x_1,x_2)$, we
consider the limiting case that one point lies at the horizon, $x_1 \rightarrow 1$, and the point where the response is measured is far away, $x_2 \rightarrow 0$. For this purpose, we can 
use the approximate solutions (\ref{asymptLsgSL}).
From the asymptotic form of $h_2^{(m)}(x_2\rightarrow 0)$ in 
Eq.\ (\ref{asymptLsgSL}), we see that the arc at infinity does not contribute to the Cauchy integral over $m^2$
in Eq.\ (\ref{eq413}).
We obtain for the retarded Green's function:
\begin{align}
G^R_{11,11}(1, x\rightarrow 0^+) & = N \sum_{n=1}^\infty m_n^2\, 
              e^{-|m_n| / x} \,  x^{2+|m|/2}\,\left[ 1 + \mathcal{O}(x) \right]\,
              \text{Res}_{m_n} \left( \frac{\mathcal{B}}{\mathcal{A}}\right)
              \notag \\
     &\sim  e^{-4.06 \, x}   + \ldots  \, , \label{sum}
\end{align}
with $N=-192 i \pi^3 \ell^4 \alpha^4 C_0 C_1 /\kappa_6^2$.
We can clearly see the exponential decay which means that transport to infinity of fluctuations of radial momentum density 
is strongly supressed. Also the dominant role of the QNM contributions $\{ m_n \}$ is clearly visible.
We have not strictly proven the convergence of the sum in Eq.\ (\ref{sum}). But although the 
residues grow with $n$, one can easily check that for small $x$
the exponential prefactors decay much faster for increasing $n$.

We would like to find the same qualitative behavior for heat transport, i.e.\ for the Green's function of energy 
density correlations, $G^R_{00,00}(x_1,x_2)$. This can easily 
be achieved, using that $h_0$ is given by the action of a first order differential operator $D_x$ on $h_2$:
\begin{align}
 h_0(x) = D_x h_2(x)  \quad \Rightarrow \quad
 G^R_{00,00}(x_1,x_2) =D_{x_1} D_{x_2} G^R_{11,11}(x_1,x_2) \, ,
\end{align}
where the exact form of the operator $D_x$ can be read off from Eq.\ (\ref{relh0h2}). We see immediately that this kind of transformation 
preserves the qualitative properties of the Green's function and in particular its exponential decay
as $ x_2 \rightarrow 0^+$.
For definiteness, we nevertheless give the result for the other Green's functions in the asymptotic limit: 
\begin{align}
 G^R_{00,00}(1,x\rightarrow 0^+) &= N \sum_{n=1}^\infty |m_n|^3\,  e^{-|m_n| /x}\, x^{1+|m_n|/2} 
       \, \left[ 1 + \mathcal{O}(x) \right]\,
       \text{Res}_{m_n}\left( \frac{\mathcal{B}}{\mathcal{A}} \right) \, , \\
 G^R_{22,22}(1,x\rightarrow 0^+) &= N \sum_{n=1}^\infty \frac{|m_n|^3}{2}\,  
       e^{-|m_n| /x}\, x^{1+|m_n|/2} \,\left[ 1 + \mathcal{O}(x) \right]
       \, \text{Res}_{m_n}\left( \frac{\mathcal{B}}{\mathcal{A}} \right) \, .
\end{align}

The exponential decay of correlators of $T_0^0$
shows that heat transport due to small perturbations in energy density near the horizon
is exponentially supressed as one goes radially towards infinity. 

\section{Summary and Discussion}
\label{sec:conclusion}
In order to get a step closer towards studying QCD-like theories via gauge/gravity duality, we started by
investigating the six-dimensional AdS soliton which is completely smooth and horizon-free, but it has one compact  
dimension which allows to break supersymmetry and conformal invariance.
In the context of gauge/gravity duality, the AdS soliton serves as a toy model to study a strongly coupled 
field theory with broken supersymmetry in a confined phase. 
On the 
other hand, we generalized the AdS soliton by observing that it can be foliated along the holographic direction
with any Ricci flat metric, in particular with a Schwarzschild black hole, giving rise to the Schwarzschild soliton string. 
The Schwarzschild soliton string serves as a 
toy model to understand the field theory dual to the AdS soliton on a non-trivial background spacetime.

By calculating the correlator $\langle T_2^1(x_1) T_2^1(x_2) \rangle$ via gauge/gravity duality, we have shown that momentum and energy diffusion 
to infinity is exponentially supressed in the quantum theory on the AdS soliton boundary.
This behavior originates from the fact that the QN frequencies appear as the poles of the momentum space Green's function, such 
that the position space Green's functions are dominated by these QNM contributions. 
The quantization of the modes is due to the particular geometry, which caps off smoothly at the IR floor. 
We leave it to future studies to 
elaborate on other modes of perturbations of the AdS soliton. Qualitatively, one expects very similar results,
although the analysis is more involved because the vector and scalar perturbations have more than just one non-zero component.

In the case of the Schwarzschild soliton string we found similar behavior of stress-energy correlators. 
The transport of energy and momentum density from the near horizon region towards infinity is exponentially supressed in 
the boundary field theory. The physical reason is again the particular bulk geometry which leads to the dominance of
QNM in the Green's functions. 

We found that due to the presence of another scale (the Schwarzschild radius $r_s$ of the black hole),
the Schwarzschild soliton string shows an interesting classical behavior under small perturbations.
The stability analysis of the tensor mode can be reduced to a combination of the classical 
Gregory-Laflamme instability and the propagation of a scalar field in the AdS soliton. 
Depending on the relation between the two physically relevant scales (i.e.\ the size $L_\tau$ of the compact dimension, and $r_s$), this spacetime is unstable if the Schwarzschild radius
is small compared to $L_\tau$. 
We found this instability at the level of spherically symmetric linearized tensor perturbations. 
We have also shown that the vector and scalar modes with respect to the base manifold Schw${}_4$
do not develop instabilities. 
Via the holographic duality, this instability is interpreted as a deconfinement transition in the field theory.

It is interesting to explore further holographic duals of field theory states on curved spacetimes. 
In \cite{MarolfAdSString}, the possibility of black droplets and black funnels has been 
discussed. 
The black funnel is a solution with a single connected horizon that is dual to the 
Hartle-Hawking state of a strongly coupled plasma around a black 
hole. Such solutions have recently been constructed numerically \cite{Funnels}.
Black droplets, on the other hand, have been conjectured to describe the final state of the AdS black string 
instability \cite{BlackStringInstability} and are still to be constructed explicitly.  
Understanding the endpoint of the Schwarzschild soliton string instability in terms of such solutions might
then allow for a complete survey of the deconfinement transition in a strongly coupled plasma
in terms of different bulk geometries.

\acknowledgments

It is a pleasure to thank my advisor Don Marolf for his support and guidance. 
I am also grateful to Jorge Santos for very useful discussions and substantial help with 
the numerics, and Mukund Rangamani for comments on a draft of this paper.
I thank the University of California, Santa Barbara
for their hospitality during the time when most of this work has been done. 
This research has been financially supported by funds from ETH Zurich and the University of
California, by the US NSF grant PHY-0855415, and by the German Nationial Academic Foundation.

\appendix
\section{Power Series Solution for Scalar Field in the AdS Soliton}
\label{appendix:AdSSoliton}
In this appendix, we outline how a power series ansatz 
leads to the solutions (\ref{SolitonSol1}, \ref{SolitonSol2}).

Plugging the ansatz $\phi_q^{(0)}(z) = \sum_{n=0}^\infty A_n z^n$
into Eq.\ (\ref{AdS_Soliton_Eq}), we find that the coefficients $A_0 $ and 
$A_5$ are free, while all others are given by
\begin{align}
 A_1 &= 0, \quad A_2 = -\frac{q^2}{6\alpha^2} A_0, \quad A_3 = 0, \quad A_4 = \frac{q^4}{24 \alpha^4}A_0,\notag \\
  A_{n+1} &= \frac{1}{n^2-3n-4} \left( \frac{q^2}{\alpha^2} A_{n-1} + (n-4)^2 \, A_{n-4} \right) \quad \text{ for } n\geq 5.
\end{align}
This yields the two Frobenius solutions (\ref{SolitonSol1}) and (\ref{SolitonSol2}) with exponents 0 and 5, respectively.
The coefficients $a_n$ and $b_n$ are defined by the $A_n$ by setting either $A_0=1$, $A_5=0$ or vice versa: 
\begin{align}
 \phi_{q}^{(0)} &= A_0 \left( 1 - \frac{q^2}{6\alpha^2} z^2  + \ldots \right)
    + A_5 z^5 \left( 1 +\frac{q^2}{14\alpha^2} z^2 + \ldots \right) \notag \\
    & \equiv A_0 \left( \sum_{n=0}^\infty a_n z^n \right) + A_5 \left( \sum_{n=5}^\infty b_n z^n \right) \, .
\end{align}

On the other hand, we can also make an ansatz for a power series solution near the IR floor $z=1$, i.e.\
$\phi_q^{(1)}(z) = \sum_{n=0}^\infty C_n (1-z)^n$.
This ansatz gives only one free coefficient $C_0$ and all others as proportional to it:
\begin{align}
 C_1 &= \frac{q^2}{5\alpha^2} C_0, \quad C_2 = \frac{q^4}{100 \alpha^4} C_0 , \quad 
  C_3 = \frac{1}{45} \left[ -\left(10+\frac{q^2}{\alpha^2} \right) C_1 + \left(40+ \frac{q^2}{\alpha^2}\right) C_2 \right], \notag \\
 C_4 &= \frac{1}{80} \left[ 10 C_1 - \left( 60+ \frac{q^2}{\alpha^2} \right)  C_2 + \left( 105 + \frac{q^2}{\alpha^2} \right)
   C_3 \right] \notag,\\
 C_{n+1} &= \frac{1}{5(n+1)^2} \bigg{[} \left( 15n^2 -10n + \frac{q^2}{\alpha^2} \right) C_n - \left( 20n^2 - 50n +30 +\frac{q^2}{\alpha^2}
   \right) C_{n-1} \notag \\
    &\quad + \left( 15 n^2 -65 n + 70 \right) C_{n-2} - \left(6n^2 -37 n +57 \right) C_{n-3} + (n-4)^2 C_{n-4}  \bigg{]}.
\end{align}
This yields the solution (\ref{SolitonSol3}) near $z=1$, where we defined again coefficients that are
independent of the overall scaling $C_0$: $c_n \equiv C_n/C_0$.

We find only one solution of this form near $z=1$ because the second solution does not have the form 
of a simple power series. The indicial equation at $z=1$ has zero as a double root, so the second independent
solution near the IR-floor contains a term of the form $\sim \log(1-z) \phi_q^{(1)}(z)$. 
Since this is divergent as $z\rightarrow 1$ and does not satisfy the Neumann condition, we discard this solution.


\section{Vector and Scalar Perturbations of the Schwarzschild Soliton String}
\label{appendix:vectorscalar}\setcounter{equation}{0}
We want to outline the analysis of linearized vector and scalar channel perturbations of the 
Schwarz-schild soliton string with respect to the 
base manifold $\mathfrak{B}=\text{Schw}_4$. We will show that linearized, spherically symmetric vector and scalar perturbations do not 
give rise to an instability. 

\subsection{Vector Perturbations}
Following the general methods for linearized gravitational perturbation theory as developed 
in \cite{KodamaIshibashiSeto,KodamaIshibashi}, we start with a vector harmonic on Schw${}_4$ which we write as
\begin{align}
 \mathbb{V}_\mu = \left[ V_0(\bar r,\theta), \; V_1(\bar r,\theta), \; 0,\; 0 \right] \, , \qquad
 (\hat \Delta + k_V^2)\mathbb{V}_\mu = 0 =\hat \nabla^\mu \mathbb{V}_\mu  \,,
\end{align}
where Greek indices and quantities with a hat refer to $\mathfrak{B}=\text{Schw}_4$.
Keeping the spherical symmetry, this vector harmonic does not have components on the sphere, 
and its remaining components are independent of $\phi$. 
Also, the mode is assumed to be time-independent, which amounts to considering the 
threshold mode for which an instability could occur. 
From solving the condition $\hat \nabla^\mu \mathbb{V}_\mu = 0$, one finds immediately that $V_1(\bar r,\theta)=\widetilde{v}_1(\theta)
\bar r^{-2}(1-1/\bar r)^{-1}$. We see that this solution is not regular at $\bar r=1$. Indeed, also from looking at
the defining equation for a harmonic vector on $\mathfrak{B}$, $(\hat \Delta + k_V^2) \mathbb{V}_\mu=0$, one can 
straightforwardly infer that for $k_V^2 \neq 0$, 
$V_1$ must vanish identically. The only remaining non-trivial condition from $(\hat \Delta + k_V^2) \mathbb{V}_\mu=0$ reads
\be
\frac{1}{\bar r^2}\left(\text{cot}\theta \; \partial_\theta V_0+ \partial_\theta^2 V_0\right) + \left(1-\frac{1}{\bar r}\right) 
\left(\frac{2}{\bar r} \partial_{\bar r} V_0 +  \partial_{\bar r}^2 V_0 \right) = -k_V^2 V_0 \, ,
\ee
which is just the eigenvalue equation for the Laplacian on $S^2$. 
Considering one particular mode with ``angular momentum'' $l$, we 
write $V_0(\bar r,\theta) = P_l(\cos \theta) \, v_0(\bar r)$, and obtain
\be
\left(k_V^2-\frac{l(l+1)}{\bar r^2}\right)v_0 + \left(1-\frac{1}{\bar r}\right) 
\left(\frac{2}{\bar r} v_0' +  v_0'' \right) = 0 \, . \label{radEq}
\ee

The perturbation of the Schwarzschild metric which can be built from the harmonic vector $\mathbb{V}_\mu$ can be written as:
\begin{align}
  h_{\mu\nu} = 2 a^2 H_T \mathbb{V}_{\mu\nu} \; , \quad 
  h_{A\mu} = a f_A \mathbb{V}_\mu \; , \quad h_{AB} = 0 \, , \label{PertVector}
\end{align}
with the tensor
\begin{align}
 \mathbb{V}_{\mu\nu} = - \frac{1}{2k_V} (\hat \nabla_\mu \mathbb{V}_\nu + \hat \nabla_\nu \mathbb{V}_\mu ) \, .
\end{align}
It turns out to be useful to formulate the 
problem in terms of the variables which are introduced in \cite{KodamaIshibashiSeto} for the gauge invariant
master equation formalism.
A gauge transformation adapted to the symmetry of the vector mode is described by a gauge vector 
$\xi_a=(\xi_\mu\, , \xi_A)$ with $\xi_\mu=a L(x^A) \mathbb{V}_\mu$, $\xi_A=0$. A gauge invariant 
variable is then given by
\begin{align}
 F_A &= f_A + \frac{a}{k_V} \nabla_A H_T \, ,
\end{align}
where $\nabla_A$ is the covariant derivative with respect to the two-dimensional orbit space. Fixing the gauge with
\be
\xi_\mu = - \frac{\ell^2 \alpha^2}{k_V z^2} H_T \left[ V_0(\bar r,\theta),0,0,0 \right] \, ,
\ee 
the variable $H_T$ is effectively set to zero and the perturbation is now TTF. 
According to Eq.\ (\ref{linEFE2}), the linearized Einstein equations reduce for a TTF perturbation to $\Delta_L h_{ab}=0$. From
this, one can immediately see that $F_y=0$ such that the Lichnerowicz equation
reduces to a radial equation in $\bar r$, and an equation for $F_z(z)$.
The radial equation is exactly the same as (\ref{radEq}), and the equation for $z$ reads
\be
\left[ \frac{4}{z^2} F_z(z) + \frac{2}{z} F_z'(z) - F_z''(z) \right] + \frac{k_V^2}{\alpha^2} \frac{1}{(1-z^5)} F_z(z) = 0 \, .
\ee
This equation with regularity boundary conditions allows for discrete modes with $k_V^2/\alpha^2 < 0$, very similar to what we found for tensor modes. Since $\alpha^2$ can by definition not be negative,
the question is: Does Eq.\ (\ref{radEq}) have regular solutions with negative eigenvalues
$k_V^2<0$? If so, then they would correspond to unstable 
vector modes. However, applying a finite differences algorithm, we find that independent of what
value $l\in\{0,1,2,\ldots\}$ takes, the equation does not have any eigenvalues $k_V^2<0$ that would correspond to an instability. 

\subsection{Scalar Perturbations}
The scalar harmonics $\mathbb{S}$ are defined by $(\hat \Delta + k_S^2 ) \mathbb{S}=0$.
Assuming that our perturbation does not break the symmetry of the underlying 2-sphere, i.e. $\mathbb{S}$ is an eigenmode of the spherical Laplacian and can thus be written as
$\mathbb{S}= P_l(\cos \theta) s(\bar r)$, the defining equation for $\mathbb{S}$ becomes
\begin{align}
 \left(k_S^2 -\frac{l(l+1)}{\bar r^2}\right)s+\left(\frac{2}{\bar r}-\frac{1}{\bar r^2}\right) s'+\left(1-\frac{1}{\bar r}\right)s'' = 0 \, . \label{scalareigen}
\end{align}
We can again find the allowed eigenvalues $k_S^2$ numerically.
As it was the case for the vector modes, we only find eigenvalues $k_S^2 \geq 0$. A comparison with vector modes thus
already stronlgy suggests that there is no instability. However, for definiteness, we complete the perturbative
analysis in the following paragraphs. 

We will show that all eigenvalues of the perturbation equations satisfiy $k_S^2/\alpha^2 \leq 0$ which is not 
compatible with the fact that Eq.\ (\ref{scalareigen}) did not yield any negative eigenvalues $k_S^2<0$. 
Therefore the only consistent eigenvalue of the full problem 
will be $k_S^2=0$ such that there is no threshold mode and no instability.

As in \cite{KodamaIshibashiSeto}, we construct the following vectors and tensors build out of 
the scalar harmonic $\mathbb{S}$: 
\begin{align}
  \mathbb{S}_\mu = -\frac{1}{k_S} \hat \nabla_\mu \mathbb{S} \; , \quad \mathbb{S}_{\mu\nu} = \frac{1}{k_S^2} \hat \nabla_\mu 
   \hat \nabla_\nu \mathbb{S} + \frac{1}{4} \hat g_{\mu\nu} \mathbb{S} \, ,
\end{align}
where $\hat{g}$ is the four-dimensional Schwarzschild metric.
From these building blocks we can construct a symmetry adapted scalar perturbation:
\begin{align}
  h_{\mu\nu} = 2a^2 \left(H_L(z) \,\hat g_{\mu\nu} 
  \mathbb{S}   + H_T(z) \mathbb{S}_{\mu\nu} \right) \; , \quad
  h_{A\mu} = a f_A(z) \mathbb{S}_\mu \; , \quad h_{AB} = f_{AB}(z) \mathbb{S} \, .
\end{align}
We proceed as in the case of vector harmonics. First, we rewrite the degrees of freedom of the perturbation $h_{ab}$ in 
terms of the following (gauge invariant) variables:
\begin{align}
 F &:= H_L + \frac{1}{d-2} H_T + \frac{1}{a} \nabla^A \bar r  X_A \, ,\\
 F_{AB}&:= f_{AB} + \nabla_{A} X_B + \nabla_B X_A \, ,\\
 &\text{where } X_A := \frac{a}{k_S} \left( f_A + \frac{a}{k_S} \nabla_A H_T \right) \, .
\end{align}
We fix the gauge by using the symmetry adapted gauge vector
\begin{align}
 \xi &= - \frac{\ell^2\alpha^2}{k_S z^2} \left[ \Lambda_1(z) \mathbb{S}_\mu, \; \Lambda_2(z) \mathbb{S}, 
   \; \Lambda_3(z) \mathbb{S} \right] \, ,  \\
   \text{with}&\quad \Lambda_1(z)=H_T(z), \quad\;\; \Lambda_2(z)= \frac{k_S z^2}{\ell^2\alpha^2} X_y(z), 
   \quad\;\; \Lambda_3(z)= \frac{k_S z^2}{\ell^2\alpha^2} X_z(z) \,  .
\end{align}
This yields the following simple expression for the scalar perturbation: 
\begin{align}
 h_{ab} = \bp h_{\mu\nu} & 0 & 0 \\ \;\;\; 0 \;\;\; & F_{yy}\, \mathbb{S} & F_{yz} \,\mathbb{S} \\ \;\;\; 0 \;\;\; & F_{yz}\, \mathbb{S} & F_{zz}\, \mathbb{S} \ep \; \quad \text{with} \quad
     h_{\mu\nu} = F(z) \frac{2 \ell^2 \alpha^2}{z^2} \mathbb{S} \, \hat{g}_{\mu\nu} \, .
\end{align} 
We use this perturbation and study the resulting Einstein field equations (\ref{linEFE2}).
This yields a set of equations for $F_{AB}(z)$ and $F(z)$. After some algebraic operations, one finds that these equations 
can be decoupled: 
\begin{align}
\frac{k_S^2}{\alpha^2}\bigg{(}\frac{k_S^2}{\alpha^2}z^2&-2 \left(1+4 z^5\right)\bigg{)} F(z)+\left(z \left(4+z^5\right)\frac{k_S^2}{\alpha^2}-\frac{8}{z} \left(1-z^5\right)^2\right) F'(z) \notag \\
   & \qquad\qquad\qquad\qquad\qquad -z^2 \left(1-z^5\right) \left(\frac{k_S^2}{\alpha^2}-\frac{4}{z^2} \left(1-z^5\right)\right) F''(z) =0\, , \label{QEVP}\\
  F_{yy}(z) &=-2 l^2\left(1-z^5\right)\left(\frac{1}{z^2}F(z)+2 \left(\frac{k_S^2}{\alpha^2}\right)^{-1} \left(1-z^5\right) \partial_z\left(\frac{1}{z^2}F'(z)\right)\right) \, ,\\
     F_{zz}(z) &=-2 l^2\frac{1}{\left(1-z^5\right)}\left(\frac{1}{z^2}F(z)-2 \left(\frac{k_S^2}{\alpha^2}\right)^{-1} \left(1-z^5\right) \partial_z\left(\frac{1}{z^2}F'(z)\right)\right)\, , \\
     F_{yz}(z) &= 0 \, .
\end{align}
Solving these equations is slightly more complicated than in the vector and tensor sectors due to the fact that 
the eigenvalue $k_S^2/\alpha^2$ appears \textit{quadratically} in Eq.\ (\ref{QEVP}).
We solve Eq.\ (\ref{QEVP}) directly by ``shooting`` the 
eigenvalue such that the regularity boundary conditions are satisfied. 
We find that (just like in the case of vector and tensor perturbations) there are no positive eigenvalues $k_S^2/\alpha^2 >0$. Instead, 
there are infinitely many discrete negative eigenvalues in addition to the eigenvalue $0$. 
Due to the lack of a compatible negative eigenvalue of Eq.\ (\ref{scalareigen}), we conclude that at the level of 
linearized perturbations which have
the previously assumed form in the scalar sector with respect to Schw${}_4$, there are no unstable modes.

\end{document}